\documentclass{llncs}
\usepackage{frameit,amsmath}
\usepackage{multirow}

\usepackage{times}

\usepackage{moresize,latexsym}
\usepackage{enumitem}

\usepackage{array}
\newcolumntype{H}{>{\setbox0=\hbox\bgroup}c<{\egroup}@{}}

\usepackage[colorinlistoftodos,prependcaption,textsize=tiny]{todonotes}
\usepackage{listings}

\usepackage[linesnumbered,ruled,vlined]{algorithm2e}

\newcommand{\lazy}{lazy initialization}
\newcommand{\pc}{\D}

\newcommand{\figfont}{\small}
\newcommand{\ok}{\emph{repOK}}

\newcommand{\seppred}[2]{\ensuremath{#1(#2)}}
\newcommand{\emp}{\btt{emp}}
\def\sep{\ensuremath{*}}
\newcommand{\pto}{{\scriptsize\ensuremath{\mapsto}}}
\newcommand{\nil}{\btt{null}}

\newcommand{\code}[1]{{\small {\ensuremath{\tt #1}}}}
\newcommand{\form}[1]{\ensuremath{#1}}
\newcommand{\btt}[1]{{\ensuremath{\tt #1}}}

\def\bool{\code{bool}}
\def\int{\code{int}}

\newcommand{\setvars}[1]{\ensuremath{\bar{#1}}}
\newcommand{\constr}{\ensuremath{\Phi}}
\def\D{\Delta}
\newcommand{\heap}{\ensuremath{\kappa}}
\newcommand{\atom}{\alpha}
\newcommand{\sepnodeF}[3]{\ensuremath{{#1}{\pto}#2(#3)}}
\newcommand{\seppredF}[2]{\ensuremath{#1(#2)}}
\newcommand{\hide}[1]{}
\def\a{a}
\def\true{\code{true}}
\def\false{\code{false}}
\newcommand{\myit}[1]{\textit{#1}}
\newcommand{\PName}{\mathcal{P}}
\def\Dns{\myit{Node}}
\def\Flds{\myit{Fields}}
\def\Store{\myit{Heaps}}
\def\Stack{\myit{Stacks}}
\newcommand{\force}{\ensuremath{\models}}
\newcommand{\sheaps}{\ensuremath{h}}
\newcommand{\sstack}{\ensuremath{s}}

\def\Locations{\myit{Loc}}

\def\Val{\myit{Val}}
\def\Var{\myit{Var}}
\newcommand{\defsym}{\ensuremath{\overset{\text{\scriptsize{def}}}{=}}}
\def\FV{\myit{FV}}
\newcommand{\pure}{\ensuremath{\pi}}
\newcommand{\subst}[2]{\ensuremath{[#1 {/} #2]}}

\newcommand{\yields}{\leadsto}
\newcommand{\myif}[3]{\code{if}~#1~\code{then}~#2~\code{else}~#3}

\newcommand{\rulen}[1]{\ensuremath{{\bf \scriptstyle [#1]}}}
\newcommand{\hlr}[3]{\ensuremath{\rulen{#1}\frac{
#2
}{#3}}}
\newcommand{\hquad}[2]{\ensuremath{#1 ~{\yields}~ #2}}
\newcommand{\cenum}[2]{\ensuremath{\code{enum}(#1,#2)}}
\def\solver{\form{\code{S2SAT_{SL}}}\,}
\newcommand{\sctx}[5]{\ensuremath{\langle#1,#2,#3,#4,#5\rangle}}
\newcommand{\sectx}[5]{\ensuremath{\langle#1,#2,#3,#4,#5\rangle}}
\newcommand{\evale}[4]{\ensuremath{#1,#2 ~{\vdash}~ #3 \Downarrow#4}}
\newcommand{\evalse}[3]{\ensuremath{#1 ~{\vdash}~ #2 \Downarrow#3}}

\newcommand{\projectp}[2]{\ensuremath{{\Pi}(#1,#2)}}

\graphicspath{{figs/}}
\definecolor{mygreen}{rgb}{0,0.6,0}
\definecolor{mygray}{rgb}{0.5,0.5,0.5}
\definecolor{mymauve}{rgb}{0.58,0,0.82}

\usepackage[firstpage]{draftwatermark}
\SetWatermarkText{\hspace*{105mm}\raisebox{148mm}{\includegraphics[scale=0.11]{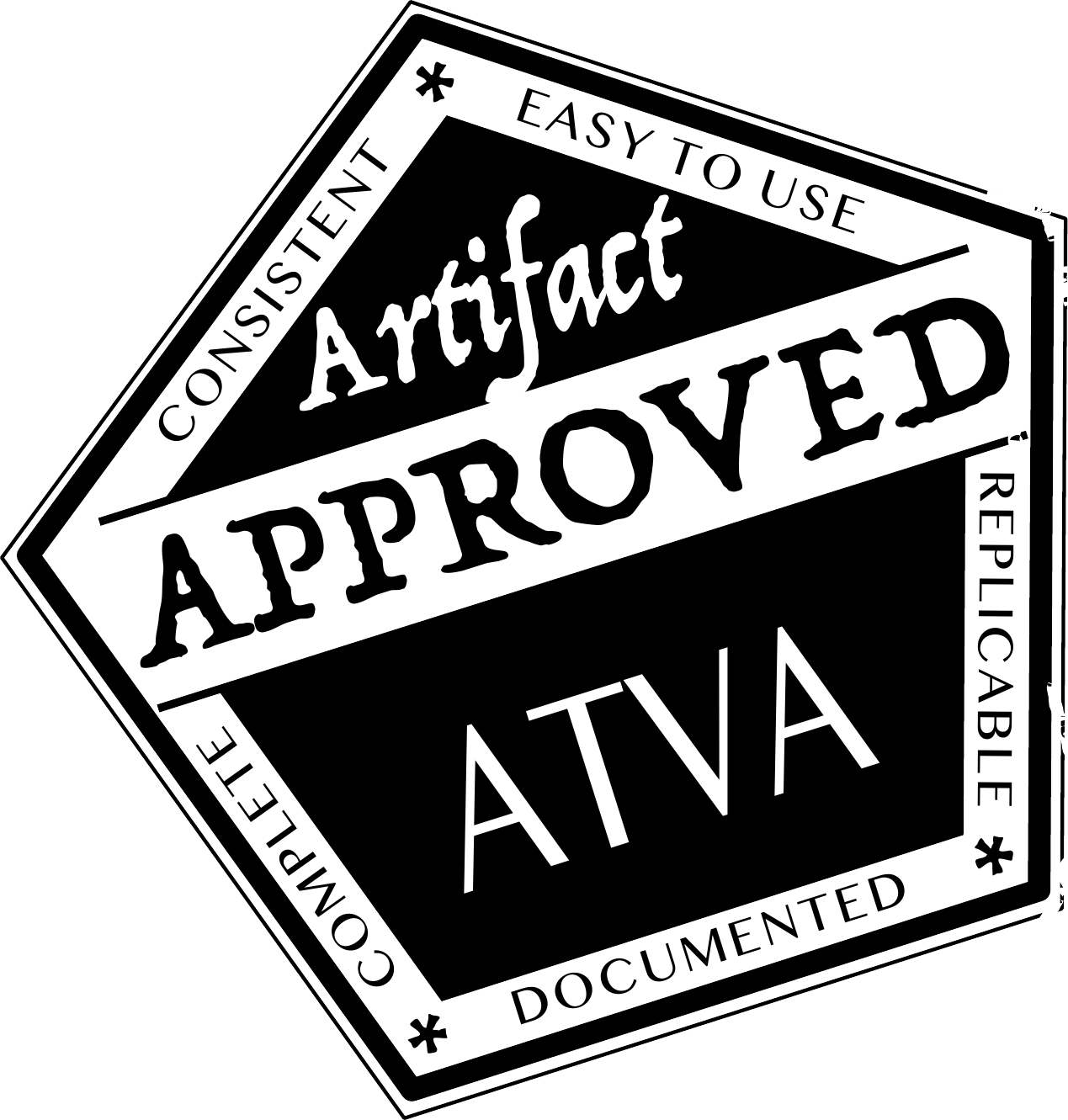}}}
\SetWatermarkAngle{0}

\begin{document}

\title{Enhancing Symbolic Execution of Heap-based Programs with Separation Logic for Test Input Generation}

\author{Long H. Pham\inst{1}\thanks{Corresponding author. Email: longph1989@gmail.com} \and Quang Loc Le\inst{2}  \and Quoc-Sang Phan\inst{3}  \and \\ Jun Sun\inst{4}  \and Shengchao Qin\inst{2}}
\institute{Singapore University of Technology and Design, Singapore \and School of Computing \& Digital Technologies, Teesside University, UK
\and {Synopsys, Inc., USA}
\and {Singapore Management University, Singapore}
}

\maketitle

\begin{abstract}

Symbolic execution is a well established 
method for test input generation.
Despite of having achieved tremendous success over numerical
domains, existing symbolic execution techniques
for heap-based programs
are limited due to the lack of
a succinct and precise description for symbolic values
over  unbounded heaps.
In this work, we present a
new symbolic execution method for
heap-based programs based on separation logic.
The essence of our proposal is 
 context-sensitive lazy initialization, a novel approach for efficient test input generation.
Our approach differs from existing approaches in two ways. Firstly, our approach is based on separation logic, which allows us to precisely capture preconditions of heap-based programs so that we avoid generating invalid test inputs. Secondly, we generate only fully initialized test inputs, which are more useful in practice compared to those partially initialized test inputs generated by the state-of-the-art tools.
We have implemented our approach as a tool, called Java StarFinder, and evaluated it on a set
of programs with complex heap inputs. The results show that our approach significantly reduces the number of invalid test inputs and improves the test coverage.

\end{abstract}

\section{Introduction} \label{sec.intro}
Symbolic execution~\cite{King:1976:SEP:360248.360252} is getting momentum thanks to its capability of discovering deep bugs. It is increasingly used not only in academic settings but also in industry, such as in Microsoft, NASA, IBM and Fujitsu~\cite{Cadar:2011:SES:1985793.1985995}.
Despite having achieved tremendous success
, symbolic execution has limited impact on testing programs with inputs in the form of complex heap-based data structures (a.k.a.~{\em heap-based}
programs). The dominant approach to symbolic execution of heap-based programs is \lazy~\cite{Khurshid2003}, which postpones the initialization of reference
variables and fields until they are accessed.
However, 
\lazy~makes no assumption on the shapes
of the input data structures, and explicitly enumerates all possible heap objects that may bind to the inputs.
This approach has the following fundamental limitations.
Firstly, due to the lack of a succinct and precise description of the shapes of the input data structures, they often generate a large number of invalid test inputs.
Secondly, due to the enumeration of all possible heap objects that may bind to the inputs, they often worsen the path explosion problem of symbolic execution. Lastly, due to lazy initialization, the generated test inputs may be partially initialized (if some fields are never accessed) and need to be further concretized.

%

In the context of logic-based verification, the problem of specifying and reasoning about heap-based programs has been studied for nearly five decades.
The dominant approaches are based on separation logic~\cite{Ishtiaq:2001:BAL:360204.375719,Reynolds:LICS02}.
The strength of separation logic lies in its separating conjunction operator \form{\sep}, which splits
the heap into disjoint regions or \emph{heaplets}. This enables \emph{local reasoning}, i.e., specification and reasoning are kept confined within heaplets, independent of the rest of the heap. This is in contrast to \emph{global reasoning}, i.e., 
the specification describes properties of the global heap,
which ``suffers from either limited applicability or extreme complexity, and scales
poorly to programs of even moderate size''~\cite{Reynolds:LICS02}.

Surprisingly, there has been limited effort on using separation logic to enhance symbolic execution for test input generation.
In this work, we start filling this gap.
Firstly, we propose a novel method for symbolic execution of heap-based programs based on separation logic.
In particular, we adopt a logic that combines separation logic with existentially quantified variables,
inductive predicates and arithmetic which allows us to encode path conditions effectively in heap-based programs.
Secondly, we enhance our method with
{\em context-sensitive} lazy initialization, i.e., we use preconditions written in separation logic to guide
 the search, and only explore the states that are reachable when the inputs satisfy the preconditions.
As a result, all generated test inputs are valid.
In summary, we make the following main contributions.
\begin{enumerate}[leftmargin=4mm]
\item We develop a symbolic execution engine for heap-based programs based on separation logic.
\item For efficiency, we present context-sensitive
lazy initialization with a {\em least fixed point} analysis
 to generate
valid test inputs during symbolic execution.
\item  We have implemented the proposed approach as a tool, called Java StarFinder\footnote{\url{https://github.com/star-finder/jpf-star}}, 
 built on top of Symbolic PathFinder~\cite{Pasareanu:2013:ASE}, to generate test inputs for Java bytecode.
\item We have evaluated our tool on a set of Java programs including complex
and mutable data structures.
All generated test inputs are valid and we achieve 98.98\% branch coverage on average.
\end{enumerate}

The rest of this paper is organized as follows. 
Section \ref{sec.motivate.dll} presents some background
and illustrates our proposal
via an example.
Section \ref{sec.prog} describes the syntax of our
core language and its operational semantics.
Our first contribution, a symbolic execution engine based on separation logic,
is presented in Section \ref{sec.se}.
Our second contribution, the context-sensitive
lazy initialization, is shown in Section \ref{sec.test}.
We present our implementation and evaluation
in Section \ref{sec.impl}.
Section~\ref{sec.related}
presents related work.
Finally, we conclude and discuss future works in
Section \ref{sec.concl}.

\section{Motivation and Illustration} \label{sec.motivate.dll}
\begin{figure}[t]
\centering
\begin{minipage}[h]{.8\linewidth}
\begin{lstlisting}
node add(node x, node y) {
  node dummyHead = new node(0, null); node z = dummyHead;
  while (x != null) {  |\label{while1}|
    z.next = new node(x.val + y.val, null);
    x = x.next; |\label{x1}| y = y.next; |\label{y1}| z = z.next;
  }
  return dummyHead.next;
}
\end{lstlisting}
\end{minipage}
\caption{Adding two numbers represented by linked lists}
\label{src:unzip}
\end{figure}

In the following, we illustrate how our approach works with an example. Consider a program that represents a big non-negative integer in the form of a
singly linked list, i.e., each node of the lists contains a single digit of the number. Suppose we want to generate test inputs for the method \form{\code{add}} shown in Fig.~\ref{src:unzip}, which computes the sum of two numbers in this representation.

This method is designed to take two parameters \form{\code{x}} and \form{\code{y}} satisfying the following preconditions: (1) all the digits of \form{\code{x}} and \form{\code{y}} are less than 5, and thus there is no carry; (2) \form{\code{x}} and \form{\code{y}} have the same number of digits. Condition (1) is a simple numerical constraint which can be handled by existing symbolic execution engines. We thus leave it out of the discussion for the sake of simplicity. To capture condition (2), we define an 
inductive predicate \form{\code{pre}} based on our fragment of
 separation logic
 as follows.
%
%
{\figfont
\[
\begin{array}{l}
\seppred{\code{pred~pre}}{a,b} ~{\equiv}~ (\emp \wedge a=\nil \wedge b=\nil) \\
\qquad \vee~ (\exists n_1, n_2.~\sepnodeF{a}{\code{node}}{\_,n_1} \sep
 \sepnodeF{b}{\code{node}}{\_,n_2} \sep
 \seppred{\code{pre}}{n_1,n_2} )
\end{array}
\]
}
%
Predicate {\emp} means the heap is empty; predicate
\form{\sepnodeF{a}{\code{node}}{\_,n_1}} states $a$ points to an allocated object; and $\sep$ is the separating conjunction operator in separation logic.
 The data type $\code{node}$ corresponds to the class $\code{node}$ in the program, which has two instance fields: $\code{val}$ containing the digit, and $\code{next}$ referencing to another $\code{node}$ object. The wildcard ``$\_$'' is used 
to indicate a ``don't care" value.

Intuitively, a linked list is recursively defined as a head points-to predicate with its next field pointing to a sublist.
In the base case of the definition, the heap is empty, and both parameters $a$ and $b$ are $\nil$. In the recursive case, \form{\sepnodeF{a}{\code{node}}{\_,n_1}} signifies that $a$ points to an allocated object composed of a certain value (represented by ``$\_$'') and its $\code{next}$ field $n_1$. Similarly, $b$ points to an allocated object with its $\code{next}$ field  $n_2$. Furthermore, $n_1$ and $n_2$, i.e., the sublists of $a$ and $b$ respectively, satisfy \form{\code{pre}} as well.

In this definition, the separating conjunction operator {\sep} splits the global heap into three heaplets. The first two heaplets contain the $\code{node}$ objects referenced respectively by $a$ and $b$, 
and the third one contains the sublists. This separation enforces that $a$ and $b$ refer to two distinct objects and their sublists are disjoint too.
Since 
 $a$ and $b$ must be either both {\nil} or both not ${\nil}$, and likewise for the objects in their sublists, 
 $a$ and $b$ have the same length.


To generate test inputs, we perform symbolic execution of method \form{\code{add}(\code{x},\code{y})} with precondition \form{\seppredF{\code{pre}}{\code{x},\code{y}}}. 
In the proposed symbolic execution, path conditions
are formulae in the fragment of
separation logic with inductive predicates and arithmetic.
Reference variables are initialized by values obtained
from a procedure, called \form{\code{enum}}.
Initially, \form{\code{x}} and \form{\code{y}} are initialized to symbolic (stack) values \form{X} and \form{Y} respectively, and the path condition $\pc$ is initialized to \form{\seppredF{\code{pre}}{X,Y}}.
 When variable \form{\code{x}} is first 
accessed
at line~\ref{while1}, 
our engine, through procedure \form{\code{enum}}, examines precondition \form{\seppredF{\code{pre}}{X,Y}}\footnote{In all path conditions in this example we only show
the constraints over those variables which are relevant to the inputs \form{\code{x}} and \form{\code{y}};
the constraints over  local variables \form{\code{z}} and \form{\code{dummyHead}} are separated from \form{\code{x}},
\form{\code{y}} and thus are omitted for simplicity.} for possible heap values for \form{\code{x}}.
Procedure \form{\code{enum}} gets possible values
through the least fixed point analysis with procedure \code{LFP}.
Procedure \code{LFP} unfolds predicate \form{\code{pre}}
until the set of values
reaches a fixed point.
In this example, procedure \code{LFP} only needs to unfold predicate \form{\code{pre}} once
and reach the fixed point with two formulae corresponding to two disjuncts in the definition of predicate
\form{\code{pre}}.
Then the engine substitutes the predicate \form{\seppredF{\code{pre}}{X,Y}} in the $\pc$
by these two formulae (with \form{\alpha}-renaming, i.e., substitutions of formal/actual parameters and of bound variables to avoid name collisions) to obtain two non-deterministic choices, and symbolic execution case splits.
 It first explores the path corresponding to the base case
 and hence the constraint over \form{X}
and \form{Y} in $\pc$ becomes $X {=} \nil \wedge Y{=} \nil$.
 We use the constraint solver {\solver} \cite{Le:CAV:2016,Tatsuta:APLAS:2016,DBLP:conf/cav/LeT0C17} to check that $\pc$ is satisfiable.
 There is no further case splitting in this path, and we have a test input where \form{\code{x}} and \form{\code{y}} are both \form{\nil}.

After exploring the base case, our symbolic executor
explores the path corresponding to the recursive case. The updated path condition $\pc$ over \form{X}
and \form{Y} is: {\figfont$$ \exists n_1,n_2 .~\sepnodeF{X}{\code{node}}{\_,n_1} \sep \sepnodeF{Y}{\code{node}}{\_,n_2}
~\textcolor{blue}{ \sep}~ \textcolor{blue}{ \code{pre}(n_1,n_2)}$$}
Executing the body of the loop, at line~\ref{x1}, \form{\code{x.next}} is dereferenced; hence $n_1$ is accessed. Since $n_1$ is constrained by \form{\code{pre}}, our engine again tries two possible values for $n_1$. 
For the base case, the path condition $\pc$ over \form{X},
 \form{Y}, \form{n_1} and \form{n_2} is:
{\figfont$$\exists n_1,n_2 .~\sepnodeF{X}{\code{node}}{\_,n_1} \sep \sepnodeF{Y}{\code{node}}{\_,n_2} ~\textcolor{blue}{\wedge~ n_1{=}\nil \wedge n_2 {=} \nil}$$}
Then, $n_2$ is accessed via \form{\code{y.next}}. Since it has been assigned to {\nil} already, there is no case splitting. $n_1{=}\nil$ violates the looping condition so symbolic execution finishes exploring the path and backtracks.
We obtain
a test input where \form{\code{x}} and \form{\code{y}} both have one digit.
Likewise, we generate test inputs where \form{\code{x}} and \form{\code{y}} both have two digits, three digits and so on. Note that we put a bound on the number of unfolding for loops. 

In contrast to ours which always generates valid test inputs,
the existing lazy approaches \cite{Visser:ISSTA:2004,Braione:2015:SEP,Braione:2016:JSE:2950290.2983940}
would generate invalid test inputs such as
(i) \form{\code{x}} and \form{\code{y}} have different number of digits;
or (ii) \form{\code{x}} and \form{\code{y}} are aliasing;
or (iii) \form{\code{x}} (or \form{\code{y}}) is a cyclic linked list.

\section{A Core Language} \label{sec.prog}
\begin{figure*}[t]
\centering
{\ssmall
\begin{minipage}{.98\textwidth}
\begin{frameit}
\[
\hlr{CONST}{
}
{\evale{\sheaps}{\sstack}{k}{k}}
\qquad
\hlr{VAR}{
}
{\evale{\sheaps}{\sstack}{v}{\sstack(v)}}
\qquad
\hlr{NULL}{
}
{\evale{\sheaps}{\sstack}{\nil}{\nil}}
\]
\[
\hlr{UNOP}{
\evale{\sheaps}{\sstack}{e_1}{k_1} \quad
k' =  ~op_{u}~ k_1
}
{\evale{\sheaps}{\sstack}{op_{u}~ e_1}{k'}}
\qquad
\hlr{BINOP}{
\evale{\sheaps}{\sstack}{e_1}{k_1} \quad
\evale{\sheaps}{\sstack}{e_2}{k_2} \quad
k' = k_1 ~op_{b}~ k_2
}
{\evale{\sheaps}{\sstack}{e_1 ~op_{b}~ e_2}{k'}}
\]
\[
\hlr{LOAD}{
\evale{\sheaps}{\sstack}{v}{k_1} \quad
 r = \sheaps(k_1) \quad
k_2 = r(\code{Type}(v), f_i)
}
{\evale{\sheaps}{\sstack}{v.f_i}{k_2}}
\quad
\hlr{FREE}{l= \sstack(v) \quad
\sheaps' = \sheaps \setminus \{l \mapsto \_\}
\quad \iota = \Sigma(pc+1)
}
{\hquad{\sctx{\Sigma}{\sheaps}{\sstack}{pc}{\btt{free}~v}}{\sctx{\Sigma}{\sheaps'}{\sstack}{pc+1}{\iota}}}
\]
\[
\hlr{ASSIGN}{
\evale{\sheaps}{\sstack}{e}{k} \quad
\sstack' = \sstack[v \leftarrow k]
\quad
\iota = \Sigma(pc+1)
}
{\hquad{\sctx{\Sigma}{\sheaps}{\sstack}{pc}{v := e}}
{\sctx{\Sigma}{\sheaps}{\sstack'}{pc+1}{\iota}}}
\]
\[
\hlr{NEW}{
\begin{array}{c}
\code{fresh{-}map}~ r' \quad r'(c,f_i) = \sstack(v_i) ~\forall i {\in} \{1..n \} \quad
\code{fresh}~ l' \\
\sheaps' = \sheaps[l' \leftarrow r'] \quad \sstack' = \sstack[v \leftarrow l']
\quad \iota = \Sigma(pc+1)
\end{array}
}
{\hquad{\sctx{\Sigma}{\sheaps}{\sstack}{pc}{v~ {:=}~\code{new}~c(v_1,...,v_n)}}{\sctx{\Sigma}{\sheaps'}{\sstack'}{pc+1}{\iota}}}
\]
\[
\hlr{STORE}{
\begin{array}{c}
\evale{\sheaps}{\sstack}{v}{k_1}
\quad \evale{\sheaps}{\sstack}{e}{k_2} \quad
 r = \sheaps(k_1) \\
r'= r[(\code{Type}(v), f_i) \leftarrow k_2] \quad
\sheaps' = \sheaps[k_1 \leftarrow r']
\quad \iota = \Sigma(pc+1)
\end{array}
}
{\hquad{\sctx{\Sigma}{\sheaps}{\sstack}{pc}{v.f_i := e}}{\sctx{\Sigma}{\sheaps'}{\sstack}{pc+1}{\iota}}}
\]
\[
\hlr{GOTO}{
\evale{\sheaps}{\sstack}{e}{k} \quad
\iota = \Sigma(k)
}
{\hquad{\sctx{\Sigma}{\sheaps}{\sstack}{pc}{\code{goto} ~e}}
{\sctx{\Sigma}{\sheaps}{\sstack}{k}{\iota}}}
\quad
\hlr{ASSERT}{
\evale{\sheaps}{\sstack}{e}{\true}
\quad \iota = \Sigma(pc+1)
}
{\hquad{\sctx{\Sigma}{\sheaps}{\sstack}{pc}{\code{assert}(e)}}{\sctx{\Sigma}{\sheaps}{\sstack}{pc+1}{\iota}}}
\]
\[
\hlr{TCOND}{
\evale{\sheaps}{\sstack}{e_0}{\true} \quad
\evale{\sheaps}{\sstack}{e_1}{k_1} \quad
\iota = \Sigma(k_1)
}
{\hquad{\sctx{\Sigma}{\sheaps}{\sstack}{pc}{\myif{e_0}{\code{goto} ~e_1}{\code{goto} ~e_2}}}
{\sctx{\Sigma}{\sheaps}{\sstack}{k_1}{\iota}}}
\]
\[
\hlr{FCOND}{
\evale{\sheaps}{\sstack}{e_0}{\false} \quad
\evale{\sheaps}{\sstack}{e_2}{k_2} \quad
\iota = \Sigma(k_2)
}
{\hquad{\sctx{\Sigma}{\sheaps}{\sstack}{pc}{\myif{e_0}{\code{goto} ~e_1}{\code{goto} ~e_2}}}
{\sctx{\Sigma}{\sheaps}{\sstack}{k_2}{\iota}}}
\]
\end{frameit}
\end{minipage}
}
\caption{Operational semantics of the core language}\label{fig.sem}
\end{figure*}

In~\cite{Schwartz:2010:YEW:1849417.1849981}, Schwartz \emph{et al.} described the algorithm of symbolic execution as an extension to the run-time semantics of a general language. The language, called SimpIL, 
 is simple but ``powerful enough to express typical languages as varied as Java and assembly code''~\cite{Schwartz:2010:YEW:1849417.1849981}. In this work, we use a similar presentation to describe our new symbolic execution engine. This section introduces our core language, which is an extension of SimpIL with 
operations on
heap memory. Note that our implementation is for Java bytecode, and our approach extends to other languages.
\paragraph{Syntax}
\begin{figure}[t]
\centering
{\small
\begin{center} 
\[
\begin{array}{rcl}
\myit{datat} & ::= & \btt{data} ~c ~\{~ \myit{field;}^* ~\} \\
 \myit{field} & ::= & t~v  \qquad
 t ::= c ~|~ \tau \qquad
 \tau ::= \bool ~|~ \int \\
\form{\myit{prog} & ::= & stmt;^*}  \\
stmt & ::= &  v~{:=}~e ~|~  v{.}f_i~{:=}~e ~|~ \code{goto}~ e
~|~ \code{assert} ~e \\
& & |~ \myif{e_0}{\code{goto}~ e_1}{\code{goto}~e_2} \\
& & |~v := \btt{new}~c(v_1,..,v_n)~|~\btt{free}~v
  \\ 
e & ::= &  k ~|~ v ~|~ v{.}f_i ~|~ e_1 ~op_{b}~ e_2 ~|~
op_{u}~e_1 ~|~\nil\\
\end{array}
\]
\end{center}
}
\caption{A core intermediate language}\label{fig.syntax}
\end{figure}

The syntax of the language is defined in Fig.~\ref{fig.syntax}.
%
A program in our core language
 consists of multiple data structures and statements.
The primitive types include boolean and integer;
statements consist of
assignment, memory store, goto, assertion,
conditional goto, memory allocation, and memory deallocation;
expressions are side-effect free
and consist of typical non-heap expressions and memory load.
We use \form{\code{op_{b}}} to represent typical binary operators, e.g.,
addition, subtraction.
Similarly, \form{\code{op_{u}}} is used to represent typical
unary operators, e.g., logical negation. 
 \form{k} represents either a boolean or integer constant.
 The expressions used together with \form{\code{goto}} should not contain variables.
For the sake of simplicity, we assume the programs are in the form of static single assignments
and 
are
well-typed in the standard way.
We note that the language can be extended to handle method calls. 

\paragraph{Operational semantics}

The \emph{concrete} execution configuration of a program defined by the syntax shown in Fig.~\ref{fig.syntax} is a tuple of five components
$\sctx{\Sigma}{\sheaps}{\sstack}{pc}{\iota}$.
\form{\Sigma} is the list of program statements; \form{\sheaps} is the current memory state (i.e., the heap); \form{\sstack} records the current value of program variables (i.e., the stack); \form{pc} is the program counter; and \form{\iota} is the current statement. Among these, \form{\Sigma}, \form{\sheaps} and \form{\sstack} are mapping functions: \form{\Sigma} maps a number to a statement; \form{\sheaps} maps a memory location to its content; \form{\sstack} maps a variable to its value.

The concrete heap \form{\sheaps} of type \form{\Store} assumes a fixed finite collection {\Dns},
a fixed finite collection {\Flds},
a disjoint set {\Locations} of locations (i.e., heap addresses),
a set of non-address values {\Val}, such that \form{\nil \in {\Val}}
and {\Val} \form{\cap} {\Locations}  {=} \form{\emptyset}.
We define \form{\Store} as:
\[
\begin{array}{lcl}
{\Store}  & {\defsym} &  {\Locations} {\rightharpoonup_{fin}} ({\Dns} ~{\rightarrow}~ \Flds ~{\rightarrow}~ \Val \cup \Locations)  \\
\end{array}
\]
Further, a concrete stack \form{\sstack} is of type \form{\Stack}, defined as follows: 
\[
\begin{array}{lcl}
{\Stack} & {\defsym} &  {\Var} ~{\rightarrow}~ \Val \cup \Locations
\end{array}
\]
is a mapping from a variable to a value or a memory address.
We use \form{[x \leftarrow k]} to denote
 updating a variable \form{x} with value \form{k} for mapping functions;
for example, \form{\sstack[x \leftarrow 13]} denotes
a new stack that is the same as stack \form{\sstack} except that it maps variable \form{x} to the value $13$. 
The operational semantics of our language
is shown in Fig.~\ref{fig.sem}. The rules
are of the following form:
\[
\begin{array}{c}
\text{computation} \\
\hline
\langle \text{current state}\rangle~\yields~\langle \text{end state}\rangle
\end{array}
\]
The computation in a rule is read from the top to the bottom, the left to the right, and
are applied based on syntactic pattern-matching.
Given a statement, our engine finds a rule to execute
the computation on the top and returns
the end state in the case of success.
If no rule matches (e.g., accessing a dangling pointer),
 the execution halts.
In these rules, \form{\code{fresh}} is used as an overloading function
to return a new variable/address. Similarly,
\form{\code{fresh{-}map}} returns a new mapping and \form{\code{Type}}
returns the type of a variable.

For the evaluation of expressions, we
use
\form{\evale{\sheaps}{\sstack}{e}{ k}}
to denote the evaluation of expression
\form{e} to a value \form{ k}
in the current context \form{\sheaps}
and \form{\sstack}.
The application of these rules is also
based on pattern-matching similar to
 the application of the statements above.

For example, rule \form{\rulen{NEW}} describes the operational
semantics of the command that allocates dynamic heaps.
Firstly, it creates a new mapping \form{r'} to relate
fields of the new object to their stack values.
Next, it generates a new heap entry at the fresh address
\form{l'}. Lastly, it updates the stack value
of the variable with the heap address.

\section{Symbolic Execution}\label{sec.se}
This section presents details on symbolic execution using a separation logic-based language
to encode path conditions in heap-based programs.



\paragraph{Separation logic} \label{sec.sl}
We use separation logic~\cite{Ishtiaq:2001:BAL:360204.375719,Reynolds:LICS02}
to capture 
symbolic heaps and expressions.
Separation logic,
an extension of Hoare logic, is a
state-of-the-art assertion language
 designed for reasoning about heap-based programs.
It provides concise and precise notations for reasoning about the heap. In particular, it supports the separating conjunction operator \form{\sep} that
splits the global heap into disjoint sub-heap regions, each of which can be analysed independently.
Combined with inductive predicates, separation logic has been shown
 to capture semantics of unbounded heaps,
 loops and recursive procedures naturally
 and succinctly \cite{Le:CAV:2016,Tatsuta:APLAS:2016,DBLP:conf/cav/LeT0C17}.

\begin{figure}[tb]
\centering
{\small
 \[
\begin{array}{cl}
 \constr &::= \D ~|~ \constr_1 ~{\vee}~ \constr_2 \\
\D & ::= \exists \bar{v}{.}~(\heap\wedge\atom\wedge\phi)
 \\
 \heap &::= \emp ~|~
\sepnodeF{v}{c}{f_i{:}v_i} ~|~
\seppredF{P}{\setvars{v}}
~|~\heap_1 \sep\heap_2 \\
 \atom &::= \true \mid  v_1 {=}
v_2~ |~ v{=}\nil \mid \neg \atom
 \mid \atom_1 \wedge \atom_2 \\ 
 \phi &::= \true  \mid
 \a_1 {=} \a_2 \mid \a_1 {\leq} \a_2
  ~|~ \neg \phi
 ~|~ \phi_1\wedge\phi_2 \\
 a &::=\!k \mid v \mid k{\times}\a \mid \a_1 \!+\! \a_2 \mid - \a \\
 \textit{Pred} & ::= \seppred{\code{pred}~P_1}{\setvars{v}_1}  \equiv \constr_1;...;\seppred{\code{pred}~P_N}{\setvars{v}_N}  \equiv \constr_n
\end{array}
\]
}
\caption{Syntax of separation logic}\label{prm.spec.fig}
\end{figure}

In the following, we define the separation logic formulae used in this work to encode path conditions of heap-based programs.
A separation logic
 formula is defined by the syntax presented in Fig.~\ref{prm.spec.fig}.
We assume that $c \in {\Dns}$ is a heap node; $f_i \in {\Flds}$ is a field; and $v,v_i$ represent variables.
We notice that each kind of heap nodes \form{c} corresponds to a data structure
declared by the user using the keyword \form{\code{data}} in our core language.
 We write \form{\bar{v}} to denote a sequence of variables.
A separation logic formula is denoted as $\constr$, which can be either a symbolic heap \form{\D} or a disjunction of them.
A symbolic heap \form{\D} is an
existentially quantified conjunction of some spatial formulae {\heap},
some pointer (dis)equalities $\alpha$,
and
 some formulae in 
 arithmetic  $\phi$ \cite{Enderton200167}.
All free variables in \form{\D}, denoted by function \form{\FV({\D})},
are either program variables or implicitly universally quantified
at the outermost level.
The spatial formula {\heap} may be a separating
 conjunction (\form{\sep})
of \form{\emp} predicate, points-to predicates \form{\sepnodeF{v}{c}{f_i{:}v_i}},
and predicate applications \form{\seppredF{P}{\setvars{v}}}. 
Whenever possible, we discard \form{f_i} of the points-to predicate and use its short form as \form{\sepnodeF{v}{c}{\setvars{v}}}.
Note that \form{v_1 {\neq} v_2}
and \form{v {\neq} \nil} are short forms for \form{\neg (v_1{=}v_2)} and \form{\neg(v{=}\nil)} respectively. 
%
Each inductive predicate is defined by a disjunction \form{\constr}
using the keyword \form{\code{pred}}.
In each disjunct, we require that variables which are not formal parameters must
be existentially quantified. 
We use \form{\D\subst{v_1}{v_2}} for a
substitution of  all occurrences
of \form{v_2} in \form{\D} to \form{v_1}.

\paragraph{Symbolic execution} \label{sec.prog2}
\begin{figure*}[t]
\begin{center}
{\ssmall
\begin{minipage}{0.98\textwidth}
\begin{frameit} 
\[
\hlr{S-CONST}{
}
{\evalse{\D,{\sstack}}{k}{k}}
\quad
\hlr{S-VAR}{
}
{\evalse{\D,{\sstack}}{v}{\sstack(v)}}
\quad
\hlr{S-NULL}{
}
{\evalse{\D,{\sstack}}{\nil}{\nil}}
\]
\[
\hlr{S-UNOP}{
\evalse{\D,\sstack}{e_1}{\pure_1} \quad
}
{\evalse{\D,\sstack}{op_{u}~ e_1}{ ~op_{u}~ \pure_1}}
\quad
\hlr{S-BINOP}{
\evalse{\D,{\sstack}}{e_1}{\pure_1} \quad
\evalse{\D,{\sstack}}{e_2}{\pure_2} \quad
}
{\evalse{\D,{\sstack}}{e_1 ~op_{b}~ e_2}{\pure_1 ~op_{b}~ \pure_2 }}
\]
\[
\hlr{S-LOAD}{
\begin{array}{c}
\sepnodeF{l}{c}{v_1,...,v_i,...,v_n} \in \D \quad \evalse{\D,{\sstack}}{v}{l} \quad
\evalse{\D,{\sstack}}{v_i}{\pure_i}
\end{array}
}
{\evalse{\D,{\sstack}}{v.f_i}{\pure_i}}
\]
\[
 \hlr{S-FREE}{\evalse{\D,{\sstack}}{v}{l} \qquad \iota = \Sigma(pc+1)
}
{\hquad{\sectx{\Sigma}{\exists \setvars{w}.~\sepnodeF{l}{c}{...} \sep\D}{\sstack}{pc}{\btt{free}~v}}{\sectx{\Sigma}{\exists \setvars{w}.~\D}{\sstack}{pc+1}{\iota}}}
\]
\[
\hlr{S-ASSIGN}{
 \quad \evalse{\D,{\sstack}}{e}{\pure} \quad
\sstack' = \sstack[v \leftarrow \pure]
\quad
\iota = \Sigma(pc+1)
}
{\hquad{\sectx{\Sigma}{\D}{\sstack}{pc}{v := e}}
{\sectx{\Sigma}{\D}{\sstack'}{pc+1}{\iota}}}
\]


\[
\hlr{S-NEW}{
\begin{array}{c}
\code{fresh}~ l' \quad
\D' \equiv \D \sep \sepnodeF{l'}{c}{v_1,...,v_n} \quad
\sstack' = \sstack[v \leftarrow l']
\quad \iota = \Sigma(pc{+}1)
\end{array}
}
{\hquad{\sectx{\Sigma}{\D}{\sstack}{pc}{v~ =~\code{new}~c(v_1,...,v_n)}}{\sectx{\Sigma}{\D'}{\sstack'}{pc{+}1}{\iota}}}
\]
\[
\hlr{S-STORE}{
\begin{array}{c}
\sepnodeF{l}{c}{v_1,...,v_i,...,v_n} \in \D \quad
\evalse{\D,{\sstack}}{v}{l}
\\ \evalse{\D,{\sstack}}{e}{\pure} \quad
\sstack' = \sstack[v_i \leftarrow \pure]
\quad \iota = \Sigma(pc+1) \\
\end{array}
}
{\hquad{\sectx{\Sigma}{\D}{\sstack}{pc}{v.f_i = e}}{\sectx{\Sigma}{\D}{\sstack'}{pc+1}{\iota}}}
\]
\[
\hlr{S-GOTO}{
\evalse{\D,\sstack}{e}{k} \quad
\iota = \Sigma(k)
}
{\hquad{\sectx{\Sigma}{\D}{\sstack}{pc}{\code{goto} ~e}}
{\sectx{\Sigma}{\D}{\sstack}{k}{\iota}}}
\quad
\hlr{S-ASSERT}{
\evalse{\D,{\sstack}}{e}{\pure} \quad \D' \equiv {\D}\wedge \pure
\quad \iota = \Sigma(pc{+}1)
}
{\hquad{\sectx{\Sigma}{\D}{\sstack}{pc}{\code{assert}(e)}}{\sectx{\Sigma}{\D'}{\sstack}{pc{+}1}{\iota}}}
\]
\[
\hlr{S-TCOND}{
\evalse{\D,\sstack}{e_0}{\pure_0} \quad
\evalse{\D,\sstack}{e_1}{k_1} \quad
\D'\equiv\D \wedge\pure_0 \quad
\iota = \Sigma(k_1)
}
{\hquad{\sectx{\Sigma}{\D}{\sstack}{pc}{\myif{e_0}{\code{goto} ~e_1}{\code{goto} ~e_2}}}
{\sectx{\Sigma}{\D'}{\sstack}{k_1}{\iota}}}
\]
\[
\hlr{S-FCOND}{
\evalse{\D,{\sstack}}{e_0}{\pure_0} \quad
\evalse{\D,{\sstack}}{e_2}{k_2} \quad
\D'\equiv\D \wedge\neg \pure_0 \quad
\iota = \Sigma(k_2)
}
{\hquad{\sectx{\Sigma}{\D}{\sstack}{pc}{\myif{e_0}{\code{goto} ~e_1}{\code{goto} ~e_2}}}
{\sectx{\Sigma}{\D'}{\sstack}{k_2}{\iota}}}
\]
\end{frameit}
\end{minipage}
}
\end{center}
\caption{Symbolic operational execution rules}\label{fig.se}
\end{figure*}

Recall that  the concrete execution configuration is a 5-tuple.
The \emph{symbolic} execution configuration is also
 a tuple of five components:
\form{\langle\Sigma, \D, \sstack, 
pc,\iota\rangle}
 where \form{\D}
is a path condition in the form of a separation logic formula defined above
and \form{\sstack} is used to map every variable
to a symbolic value\footnote{We use the same symbol \form{\sstack} as in
concrete setting. From the context, it should be clear as to whether we are referring to
symbolic stack or concrete stack.}.
The rest of the components are similar to those of the concrete execution configuration, 
except that symbolic values of variables
are captured in \form{\D} and \form{\sstack}. 
 We use \form{\pure} (and \form{\pure_i})
 to denote symbolic values.
Memory allocations are symbolically
 captured in the path condition \form{\D}.
A symbolic execution configuration \form{\langle\Sigma, \D, \sstack, 
pc,\iota\rangle}
is {\em infeasible} if
\form{\D}
 is unsatisfiable.
Otherwise, it is {\em feasible}.

All operational symbolic execution rules over
our language are shown in Fig.~\ref{fig.se}. 
In these rules, similar to Fig.~\ref{fig.sem} we use function \form{\code{fresh}} to return
a fresh variable. 
We illustrate the execution through rule \rulen{S-NEW}. This rule
allocates a new object of type \form{c} and assigns to variable \form{v}. 
Firstly, it generates a fresh symbolic address \form{l'}
and updates the stack to map \form{v} to this address.
Secondly, it creates new symbolic heap for \form{l'} 
by separately conjoining the current path condition with
 a new points-to predicate \form{ \sepnodeF{l'}{c}{v_1,...,v_n} }.
Lastly, it loads the next statement using the program counter.
Note that we assume all variables $v_1,...,v_n$ used in memory allocation are distinct and
each variable $v_i$ is only used to create at most one new object.
%

Rule \rulen{S-FREE} symbolically de-allocates the heaps.
To capture the de-allocated heaps for test input generation,
we keep track the corresponding points-to predicates
by storing them in 
a ``garbage'' formula. At the end of execution, those predicates are plugged into the
current path condition before being used to generate the test input.

\section{Lazy Test Input Generation}\label{sec.test}
In this section, we present the test input generation based on the symbolic execution engine we depicted in the previous sections.
The inputs of our method are a program \form{\code{prog}} in the language we defined in Section~\ref{sec.prog} and a precondition
\form{\D_{pre}} in the form of a separation logic formula defined in Section~\ref{sec.sl}.
The output is a set of fully initialized test inputs that satisfy the precondition and often achieve high test coverage.
Our method is based on lazy initialization.
The main difference between our method and previous approaches based on lazy initialization is that we generate values
of reference variables and fields in a context-sensitive manner.

Our symbolic execution engine starts with the 
 configuration:
$
\form{\langle\Sigma, \D_{pre}\sigma, \sstack_0, 
pc_0, \iota_0\rangle}
$
where \form{\sigma} is a substitution
 of input variables
 to their corresponding symbolic values, \form{\sstack_0} 
 is an initial mapping of input variables
to symbolic values,
 \form{pc_0} and \form{\iota_0} denote the first value of the program counter
 and the first statement respectively.
The engine systematically derives the strongest postcondition of
 every program path (with a bound on the number of loop unfolding), by applying the symbolic operational execution rules (i.e., 
shown in Fig.~\ref{fig.se}). 
After obtaining the strongest post-state, 
our engine invokes {\solver} to check whether or not the resultant symbolic heap is satisfiable, and 
 generate a model if it is. The model is then transformed into a test input. 

\paragraph{Context-sensitive lazy initialization} Recall that lazy initialization~\cite{Khurshid2003} leaves a reference variable or field
uninitialized until it is first accessed and then enumerates all possible valuations of the variable or field.
For instance, as shown in Figure~\ref{fig.se}, a reference variable or field can be accessed in the rules
$\rulen{S-VAR}$, $\rulen{S-LOAD}$, $\rulen{S-STORE}$, or $\rulen{S-FREE}$.
The problem is that many valuations obtained through enumeration are invalid (i.e., violating the precondition). Thus, in this work, we propose context-sensitive enumeration.
When a reference variable or field with symbolic value $v$ is accessed during symbolic execution, procedure \form{\cenum{v}{\D,\sstack}} is invoked to non-deterministically initialize $v$ with values derived from the symbolic execution context, i.e.,
 \form{\D} and \form{\sstack}.
For each value of the set, the symbolic execution engine creates a new branch for exploration.
 %
%
Procedure \form{\cenum{v}{\D,\sstack}} works based on
the following three scenarios.
\begin{enumerate}
\item If \form{\D} implies that \form{v} has previously been initialized
to either \form{\nil} (i.e., \form{\D {\Rightarrow} v{=}\nil})
or a points-to predicate (i.e., \form{\D {\Rightarrow} v{\mapsto}\_}),
 initializing it again is not necessary.
\item If \form{v} is uninitialized and there does not exist
a predicate \form{\seppredF{P_i}{\setvars{v}_i}} in \form{\D}
such that 
\form{\D \Rightarrow v \in \setvars{v}_i}  (i.e., \form{v} is not constrained by the context), $v$ is
initialized to \form{\nil}, 
 to a new points-to predicate with uninitialized fields,
or 
 to a points-to predicate in \form{\D}.
\item If \form{v} is uninitialized and there exists
some predicates
\form{\seppredF{P_i}{\setvars{v}_i}} in \form{\D}
such that 
\form{\D \Rightarrow v \in \setvars{v}_i} (i.e., \form{v} is constrained by the context), 
we substitute \form{P_i} by \form{SV_i} to instantiate \form{v}
where
 \form{SV_i} is a set of possible values
of $v$ which are consistent with the context \form{\D},
 and computed in advance
by using the least fixed point analysis (as shown below).

\end{enumerate}
%
%
\paragraph{Least fixed point analysis}

\begin{algorithm*}[t]
\figfont
%
   $SV_i \leftarrow \{P_i(\setvars{t}_i)\}$ ; ~
   $A_i \leftarrow \false$ \tcc*{i $\in$ \{1..N\}}
 \While { $\true$ }
 {
   $SV'_i \leftarrow \{\}  $ \tcc*{i $\in$ \{1..N\}}
\ForEach{$i \in \{1..N\}$}{
   \ForEach{$\D_j \in SV_i $}{
     $SV'_i \leftarrow SV'_i \cup \code{unfold}(\D_j)$ \; 
   }
    $A'_i \leftarrow  \bigvee \{\exists \setvars{w}.~\code{abs}(\projectp{\heap \wedge \atom}{\setvars{t}_i}
   , \setvars{t}_i)
  \mid (\exists \setvars{w}.~\heap \wedge \atom \wedge \phi) \in
 SV'_{i}\}$ \;
}
    \uIf{$ \forall i \in \{1..N \}.~ A'_i \Rightarrow A_i$}{
       \Return{$\bar{SV}$}  \tcc*{fixed point}
     }\uElse{
      $SV_i \leftarrow SV'_i $ ;~
      $A_i \leftarrow A'_i $ \tcc*{i $\in$ \{1..N\}}
     }
 }
 \caption{Procedure \code{LFP}}\label{algo.lfp}
\end{algorithm*}

This fixed point analysis is made use in the third case above.
We assume that
there are \form{N} predicate definitions 
\form{P_1(\setvars{t}_1)}, ..., \form{P_N(\setvars{t}_N)}
(\form{\PName} for short).
We then use the procedure \form{\code{LFP}}
 to compute
 \form{SV_1}, ..., \form{SV_N} (\form{\bar{SV}} for short)
 according to each predicate. Each \form{SV_i} stores all possible contexts
 (in the form of separation logic formulae) 
 for \form{\setvars{t}_i}
 that
could be derived from \form{P_i(\setvars{t}_i)}.
After having all \form{\bar{SV}}, in scenario 3 mentioned above,
we substitute \form{P_i(\setvars{v}_i)} with \form{SV_i[\setvars{v}_i/\setvars{t}_i]}
to get all possible values for $v$.
Because of the substitution, new variables may be introduced in \form{\D}, we update
\form{\sstack} accordingly by using the variables' names as their symbolic values.
Note that we only compute \form{\bar{SV}} once
before running the symbolic execution engine.

 The details of \form{\code{LFP}} are shown in Algorithm~\ref{algo.lfp}. \form{\code{LFP}} takes
the set of predicate definitions \form{\PName} as input.
It outputs the set of symbolic heap formulae \form{\bar{SV}} of all predicates. 
In this algorithm, each \form{A_i}, a disjunctive base formula (i.e., a formula without
any occurrence of inductive predicates), captures the abstraction
of all formulae in accordance with \form{SV_i}.
In intuition,
\form{\code{LFP}} iteratively explores each
\form{SV_i} (initialized with \form{\{P_i(\setvars{t}_i)\}} at line 1) into a set of disjoint, complete and ``smaller'' contexts.
At the same time, it computes an abstraction $A_i$
 over heap allocations 
 and (dis)equality constraints over \form{\setvars{t}_i}.
If the fixed point (at lines 8-9) is achieved
i.e., the complete set of base formulas of every predicate
has been explored, 
\form{\code{LFP}}
stops. Otherwise, it moves to the next iteration.

In particular, for the first task \form{\code{LFP}} enumerates all
possible symbolic heap locations which \form{\setvars{t}_i}
 can be assigned to.
The enumeration is performed through the function \form{\code{unfold}(\D_j)} (at line 6),
 which replaces every occurrence of inductive predicates in \form{\D_j} by
 their 
 corresponding
definitions
 with \form{\alpha}-renaming.
As a result, each disjunct in \form{SV_i} captures a new context.
%
For the second task,
after the new contexts have been derived, at line 7,
 \form{\code{LFP}} computes an abstraction on
 the set of symbolic values which every parameters \form{\setvars{t}_i} can be assigned to.
This abstraction is critical for the termination of the algorithm
and is computed by two functions. 
Intuitively, these two functions compute
constraints on parameters of each inductive predicate.
The first function $\code{abs}(\form{\heap \wedge \atom, \setvars{t}_i})$ captures the
reference values of \form{\setvars{t}_i}, while the second function
 $\projectp{\heap \wedge \atom}{\setvars{t}_i}$ captures
 (dis)equality constraints on
\form{\setvars{t}_i}.
Note that because we only want to capture heap allocations and (dis)equality constraints over \form{\setvars{t}_i},
\form{\code{LFP}} does not consider arithmetic constraints when it computes the abstraction.

 In particular, the function \form{\code{abs}} is defined as $\code{abs}(\form{\emp \wedge \atom, \setvars{v}}) = \emp \wedge \atom$.
 Otherwise,
{\figfont
\[
\begin{array}{l}
\form{\code{abs}(\sepnodeF{v}{c}{\setvars{w}} \sep \heap_1 \wedge \atom, \setvars{v})}=
\begin{cases}
 \form{ \sepnodeF{v}{c}{\setvars{w}} \sep \form{\code{abs}(\heap_1 \wedge \atom, \setvars{v}) } & \text{if } \form{v{\in}\setvars{v}  } \\
\form{ \form{\code{abs}(\heap_1 \wedge \atom, \setvars{v}) }} &  \text{otherwise }}
\end{cases}
\\
\form{\code{abs}(\form{P(\setvars{w}) \sep \heap_1 \wedge \atom, \setvars{v})}}=
\begin{cases}
 \form{ \false } & \text{if } \form{\setvars{w} \cap \setvars{v} \neq \emptyset  } \\
\form{ \form{\code{abs}(\heap_1 \wedge \atom, \setvars{v}) }} &  \text{otherwise }
\end{cases}
\end{array}
\]
}
In principle, this function retains all heap nodes allocated
by variables in \form{\setvars{v}},
maps inductive predicates with arguments in \form{\setvars{v}} to \form{\false},
and discards other constraints in \form{\heap}.

The function \projectp{\heap \wedge \atom}{\setvars{v}}
eliminates (dis)equality constraints in \form{\heap \wedge \atom}
on all variables 
 which are not in \form{\setvars{v}}.
In particular,
\form{ \projectp{\heap \wedge \true}{\setvars{v}} {=} \heap \wedge \true},
\form{ \projectp{\heap \wedge \false}{\setvars{v}} {=} \false}, and
 \form{ \projectp{\heap \wedge v_1{\not=}v_1 \wedge \atom_1}{\setvars{v}} {=} \false}.
 Otherwise,
\[
\figfont
\begin{array}{l}
 \projectp{\heap \wedge v_1{=}e \wedge \atom_1}{\setvars{v}} =
  \begin{cases}
	\projectp{\heap \wedge \atom_1[\nil/v_1]}{\setvars{v}}     &  \text{if } e{=}\nil \wedge v_1{\not\in}\setvars{v}\\
    \projectp{(\heap \wedge \atom_1)[v_2/v_1]}{\setvars{v}}       &  \text{if } e{=}v_2 \wedge v_1{\not\in}\setvars{v}\\
    \projectp{(\heap \wedge \atom_1)[v_1/v_2]}{\setvars{v}}  &  \text{if } e{=}v_2 \wedge v_2{\not\in} \setvars{v} \wedge v_1 {\in} \setvars{v}\\
   \projectp{\heap \wedge \atom_1}{\setvars{v}} \wedge v_1{=}e & \text{otherwise}
  \end{cases}
\\
 \projectp{\heap \wedge v_1{\not=}e \wedge \atom_1}{\setvars{v}} =
  \begin{cases}
    \projectp{\heap \wedge \atom_1}{\setvars{v}} \wedge v_1{\not=}e    &   \text{if }e{=}\nil \wedge v_1 {\in} \setvars{v} \vee e {=}v_2 \wedge v_1{\in}\setvars{v} \wedge v_2{\in}\setvars{v}\\
     \projectp{\heap \wedge \atom_1}{\setvars{v}}   & \text{otherwise}
  \end{cases}
\end{array}
\]
An equality \form{v_1{=}v_2} is retained if both \form{v_1} and
\form{v_2} are in \form{\setvars{v}}. Otherwise, it is eliminated
and one of the variables must be
eliminated via a substitution.
A disequality \form{v_1{\neq}v_2} is retained if both \form{v_1} and
\form{v_2} are in \form{\setvars{v}}. Otherwise, it is eliminated.
Similarly, \form{v_1{=}\nil} and \form{v_1{\not=}\nil} are retained
if \form{v_1} is in \form{\setvars{v}}. After applying the above two functions,
formulae may contain redundant variables in \form{\exists\setvars{w}}, which may be eliminated.

\paragraph{Correctness} Correctness of the proposed
enumeration method follows the correctness of the
procedure \form{\code{LFP}}. 
We argue that \form{\code{LFP}} is sound (i.e., all generated values are correct),
terminating, and
complete (i.e., 
all possible heap and (dis)equality constraints between reference
parameters in each predicate are captured at fixed point).
%

\begin{theorem}[Soundness]
If \form{\D_j \in SV_i}
and
\form{\sheaps, \sstack \force \D_j}
then
\form{\sheaps, \sstack \force P_i(\setvars{t}_i)}.
\end{theorem}

\begin{proof}
The soundness of \form{\code{LFP}} follows the correctness
of the unfolding, i.e., \form{\D_j} is derived
through the unfolding of \form{P_i(\setvars{t}_i)} and \form{P_i(\setvars{t}_i) \equiv \bigvee \{ \D_j \mid \D_j \in SV_i \}}.
Hence, \form{\D_j} is an under-approximated formula of \form{P_i(\setvars{t}_i)}.
\end{proof}

\begin{theorem}[Termination]
Suppose \form{M} be the maximal arity among the inductive predicates
\form{P_i(\setvars{t}_i)}, \form{i\in\{1..N\}}.
Then \form{\code{LFP}} runs in  $\mathcal{O}(N2^{M^2+M})$.
\end{theorem}

\begin{proof}
The complexity of \form{\code{LFP}} relies on the number of disjuncts computed
by two functions \form{\code{abs}(\form{\heap \wedge \atom, \setvars{t}_i})} and $\projectp{\heap \wedge \atom}{\setvars{t}_i}$.
It comes from the following three sub-components.
\begin{itemize}
\item As
the maximal arity of \form{\setvars{t}_i} is \form{M}, the number of (dis)equalities
among these variables
 is $\mathcal{O}({M^2})$. The number of its subsets is $\mathcal{O}({2^{M^2}})$.
\item Furthermore, the maximum number of points-to predicates
is \form{M}. 
 Hence, the number of its subsets is  $\mathcal{O}({2^{M}})$.
\item There are \form{N} predicate definitions in the system.
\end{itemize}
Hence, the implication at line 8 in Algorithm \ref{algo.lfp} holds in a finite number of iterations.
\end{proof}

\begin{theorem}[Completeness]
If \form{\sheaps, \sstack \force P_i(\setvars{t}_i)}
then \form{\exists\D_j \in SV_i} s.t.
\form{\sheaps, \sstack \force \D_j}.
Moreover, \form{SV_i} captures all possible heap and (dis)equality constraints
between reference variables in \form{\setvars{t}_i}.
\end{theorem}
\begin{proof}
The first part follows the correctness of the unfolding.
For the second part, notice that at least one of \form{A_1, ..., A_N}
gets weaker via each iteration until all of them reach fixed point.
Each \form{A_i} is derived from its according \form{SV_i} by
two functions \form{\code{abs}(\form{\heap \wedge \atom, \setvars{t}_i})} and $\projectp{\heap \wedge \atom}{\setvars{t}_i}$,
which captures all heap and (dis)equality constraints between \form{\setvars{t}_i}.

\end{proof}



\begin{example}
We demonstrate the computation of \form{SV} for predicate \form{\code{pre}(a,b)} in
Section \ref{sec.motivate.dll}.
The computation is summarized in Fig. \ref{fp.example} where \form{i}
is the number of the iteration.

\begin{figure*}[tb]
{\ssmall
\[\setlength\arraycolsep{2pt}
\begin{array}{l|l|l}
i & \begin{array}{l} SV^i \end{array} & \begin{array}{l} A^i \end{array} \\
\hline
0 & \begin{array}{l} \code{pre}(a,b) \end{array} & \begin{array}{l} \false \end{array} \\
\hline
1 &
\begin{array}{l} \emp \wedge a{=}\nil \wedge b{=}\nil \\
~\vee \exists n_1,n_2.~\sepnodeF{a}{\code{node}}{\_,n_1} \sep
 \sepnodeF{b}{\code{node}}{\_,n_2} \sep\seppredF{\code{pre}}{n_1{,}n_2}
\end{array}
& 
\begin{array}{l}
\emp\wedge a{=}\nil\wedge b{=}\nil \\~\vee \exists n_1,n_2.~\sepnodeF{a}{\code{node}}{\_, n_1} \sep \sepnodeF{b}{\code{node}}{\_, n_2}
\end{array}
\\
\hline
2 &
\begin{array}{l} \emp \wedge a{=}\nil \wedge b{=}\nil\\
 ~\vee \exists n_1,n_2.~\sepnodeF{a}{\code{node}}{\_,n_1} \sep 
 \sepnodeF{b}{\code{node}}{\_,n_2} \wedge n_1{=}\nil \wedge n_2{=}\nil \\
~\vee \exists n_1,n_2,n_3,n_4.~\sepnodeF{a}{\code{node}}{\_,n_1} \sep
 \sepnodeF{b}{\code{node}}{\_,n_2}~\sep \\
 \quad\sepnodeF{n_1}{\code{node}}{\_,n_3} \sep  \sepnodeF{n_2}{\code{node}}{\_,n_4} \sep \seppredF{\code{pre}}{n_3{,}n_4}
\end{array}
&
\begin{array}{l}
\emp\wedge a{=}\nil \wedge b{=}\nil \\~\vee 
\exists n_1,n_2.~\sepnodeF{a}{\code{node}}{\_, n_1} \sep \sepnodeF{b}{\code{node}}{\_, n_2}
\\~\vee \exists n_1,n_2.~\sepnodeF{a}{\code{node}}{\_, n_1} \sep \sepnodeF{b}{\code{node}}{\_, n_2}
\end{array} \\ 
\end{array}
\]
}
\caption{\form{\code{LFP}} for the motivating example}\label{fp.example}
\end{figure*}

Since \form{A^2 \Rightarrow A^1},
\form{\code{LFP}} stops after two iterations
and produces \form{SV^1} as the set of new contexts
for this example.
After that, the engine substitutes
\form{SV^1} into \form{\D} to obtain the two corresponding
symbolic heaps.
\end{example}

\section{Implementation and Evaluation}\label{sec.impl}

\begin{table}[t]
	\scriptsize
    \centering
	\caption{Experimental results}
	        \begin{tabular}{| l | H H H H c | c | c | c | c | c | c | c | c | c | c | c |}
				\hline
	    \multirow{2}{*}{Program} & \multicolumn{4}{H}{CSF} & \multicolumn{4}{c|}{JSF} & \multicolumn{4}{c|}{JBSE} & \multicolumn{4}{c|}{BBE} \\
				\cline{2-17}
	             & \#Tests & Cov.(\%) & \#Calls & T(s) & \#Tests & Cov.(\%) & \#Calls & T(s) & \#Tests & Cov.(\%) & NCov.(\%) & T(s) & \#Tests & Cov.(\%) & NCov.(\%) & T(s) \\
    \hline
    DLL & 75 & \textbf{100} & 40/58 & 32 & 74/74 & \textbf{100} & 325 & 49 & 121/5146 & 56 & 100 & 206 & 0/35 & 0 & 21 & 21 \\
    \hline
    AVL & 62 & \textbf{100} & 36/654 & 274 & 69/69 & \textbf{100} & 623 & 400 & 76/295 & \textbf{100} & 100 & 48 & 17/117 & 70 & 89 & 69 \\
    \hline
    RBT & 133 & \textbf{99} & 14/1106 & 2403 & 314/314 & \textbf{100} & 2070 & 2256 & 137/291 & 87 & 91 & 38 & 14/380 & 26 & 53 & 333 \\
    \hline
    SUSHI & 5 & \textbf{100} & 3/38 & 8 & 7/7 & \textbf{100} & 30 & 5 & 0/900 & 0 & 100 & 24 & 2/27 & 25 & 25 & 8 \\
    \hline
    TSAFE & 16 & \textbf{59} & 1/595 & 1190 & 5/5 & \textbf{24} & 13 & 3 & 0/32 & 0 & 5 & 10 & 0/1 & 0 & 0 & 1 \\
    \hline
    Gantt & 22 & \textbf{100} & 2/156 & 25 & 21/21 & \textbf{100} & 140 & 25 & 17/887 & 55 & 90 & 24 & 0/6 & 0 & 5 & 2 \\
    \hline
    SLL & 29 & \textbf{100} & 21/8 & 11 & 26/26 & \textbf{100} & 55 & 11 & - & - & - & - & 16/50 & 66 & 71 & 19 \\
    \hline
    Stack & 18 & \textbf{100} & 16/2 & 7 & 18/18 & \textbf{100} & 31 & 7 & - & - & - & - & 11/14 & 84 & 84 & 6 \\
    \hline
    BST & 47 & \textbf{100} & 16/33 & 14 & 182/182 & \textbf{100} & 698 & 241 & - & - & - & - & 19/260 & 69 & 86 & 131 \\
    \hline
    AAT & 46 & \textbf{99} & 21/352 & 277 & 103/103 & \textbf{100} & 1179 & 1981 & - & - & - & - & 3/166 & 6 & 43 & 111 \\
    \hline
    Tll & 6 & \textbf{100} & 2/4 & 2 & 3/3 & \textbf{100} & 11 & 2 & - & - & - & - & 1/4 & 38 & 50 & 2 \\
    \hline
    \end{tabular}
\label{tbl:sep}
\end{table}

We have implemented our approach described in previous sections into a tool, called Java StarFinder (JSF), consisting of 11569 lines of Java code. 
The architecture of JSF was briefly described in our previous work~\cite{Pham:2018:THP:3183440.3194964}.
In the following, we evaluate JSF in order to answer three  research questions (RQ). All experiments are conducted
on a laptop with  2.20GHz Intel Core i7 and 16 GB RAM.

Our experimental subjects include \emph{Singly Linked List} (SLL), \emph{Doubly Linked List} (DLL), 
 \emph{Stack}, \emph{Binary Search Tree} (BST), and \emph{Red Black Tree} (RBT) from SIR~\cite{sir}; \emph{AVL Tree} (AVL) and \emph{AA~Tree} (AAT) from Sierum/Kiasan~\cite{sireum}, the motivation example, \emph{TSAFE} project, and \emph{Gantt} project used in SUSHI~\cite{Braione:ISSTA:2017} and a data structure called \emph{Tll} 
 ~\cite{Le:CAV:2014}.
Since JSF is yet to support string, array, and object oriented features such as inheritance and polymorphism, we exclude data structures and methods which rely on these features, i.e., \emph{Disjoint Set} in Sierum/Kiasan and \emph{Google Closure} in SUSHI. Supporting these features is left for future work.
In total, our experimental subjects include a total of 74 methods, whose lines of code range from dozens to more than one thousand.


\paragraph{RQ1: Can JSF reduce invalid test inputs?}
To answer this question, we need to check whether a generated test input is valid or not. In
the benchmark programs which we collect, six data structures contain $repOK$ methods which are designed to check if the input is valid or not, i.e.,
\emph{Stack}, \emph{DLL}, 
 \emph{BST}, \emph{RBT}, \emph{AVL}, and \emph{AAT}.
In addition, the motivation example in SUSHI is based on \emph{DLL} 
 and thus we use the method  $repOK$ of \emph{DLL}
 to validate its test inputs.
We write the $repOK$ methods manually for the remaining test subjects. For \emph{SLL}
 and \emph{Tll}, we write their $repOK$ methods based on their standard definition.
For \emph{TSAFE} and \emph{Gantt}, we write their $repOK$ methods after reading the source code, i.e., the $repOK$ encodes the condition required to avoid $RuntimeException$ such as $NullPointerException$.

For each generated test input, we check its validity by passing it as arguments to the corresponding $repOK$ method~\cite{Visser:ISSTA:2004}. If $repOK$ returns $\code{true}$, the test input is deemed valid.
As a baseline, we compare JSF with JBSE~\cite{Braione:2016:JSE:2950290.2983940}, which implements the HEX approach~\cite{Braione:2015:SEP}, and the black box enumeration (BBE) approach documented in~\cite{Visser:ISSTA:2004}. We do not compare with the white box enumeration approach~\cite{Visser:ISSTA:2004} as it requires user-provided \emph{conservative} $repOK$ methods, which are missing in these benchmarks. Note that conservative $repOK$ methods are different from $repOK$ methods, and writing those methods is highly nontrivial. We do not compare our approach with SUSHI because SUSHI generates test cases in form of sequence of method calls whereas we generate test cases in form of input data structures. SUSHI and our approach are thus complementary to each other. To run JBSE, we need invariants written in HEX. We manage to find invariants for
\emph{DLL},
 \emph{RBT}, \emph{AVL}, \emph{TSAFE}, and \emph{Gantt} from~\cite{sushiexperiment}, and thus we are able to run JBSE on these subjects.
It is not clear to us how to write HEX invariants for other data structures or if HEX is expressive enough to describe them.

The experimental results are shown in Table \ref{tbl:sep}, where the first column show the name of test subjects, 
and the last three columns show the results of JSF, JBSE, and BBE respectively.
Columns \#Tests show the results in form of the number of valid test inputs over the number of generated test inputs. 
Note that because JBSE generate partial initialized test inputs, we add an additional call to $repOK$ method after the method under test to concretize their test inputs.
The results show that, as expected, every test input generated by JSF is valid.
In comparison, JBSE generates 4.65\% valid test inputs and BBE generates 7.83\% valid test inputs.
From the results, we conclude that JSF is effective in generating valid test inputs.

\paragraph{RQ2: Can JSF generate test inputs that achieve high code coverage?}
To answer this question,
we use JaCoCo \cite{jacoco} to measure the branch coverage of test inputs generated by the tools. 
The results are shown in the columns titled Cov.(\%) and NCov.(\%) in Table \ref{tbl:sep}.
The columns NCov.(\%) show the coverage of all generated test inputs,
the columns Cov.(\%) show the coverage of test inputs that satisfy $repOK$ methods.
As all test inputs generated by JSF satisfy $repOK$ methods, the result of JSF has only one column Cov.(\%).

For 73/74 methods (including auxiliary methods), JSF can achieve 100\% branch coverage (excluding infeasible branches). The only exception is method $TS\_R\_3$ in the \emph{TSAFE} project. It is because this method invokes native methods and handling native methods is beyond the capability of JSF at the moment. In general, JSF achieves 98.98\% coverage on average.
In comparison, when considering all test inputs, JBSE achieves 95.59\% coverage on average and BBE achieves 54.66\% coverage on average. Since many of these test inputs are invalid, these coverage are inflated.
When considering only test inputs that satisfy $repOK$ method, JBSE is only able to achieve 68.54\% coverage on average and BBE achieves 37.85\% coverage on average.
From these results, we conclude that JSF can generate test inputs with high branch coverage for the methods under test.

\paragraph{RQ3: Is JSF sufficiently efficient?}
To answer this question, we measure the time spent to generate test inputs for each method.
The results are shown in the columns titled T(s) in Table~\ref{tbl:sep}. From the results, JBSE and BBE are clearly faster than JSF.
That is, JBSE and BBE takes average 8.75 and 9.50 seconds respectively to handle each method, 
whereas JSF's time ranges from 1 second to half an hour, with an average of 67.29 seconds per method.
We also report the number of solver calls used by JSF.
In average, JSF needs 70 calls per method.
The main reason JSF is slower than JBSE and BBE is JSF has to solve harder path conditions
with inductive predicates.
However, the efficiency of
 JBSE and BBE comes with the tradeoff of excessive number of invalid test inputs as discussed above, whereas JSF only generates valid test inputs.
 From these results, we conclude that JSF is slower than JBSE and BBE, but still sufficiently efficient to provide higher quality results.

\section{Related Work} \label{sec.related}
This work is based on \emph{generalized symbolic execution} (GSE)~\cite{Khurshid2003}, which is the state-of-the-art way for the symbolic execution~\cite{King:1976:SEP:360248.360252} of heap-based programs. At the heart of GSE is the {\lazy} algorithm which executes programs on inputs with reference variables and fields  being \emph{uninitialized}. When a reference variable or field is first accessed, {\lazy} enumerates all possible heap objects that it can: (i)~be \form{\code{null}}, (ii)~point to a new object with all reference fields being uninitialized, or (iii)~point to any previously initialized object of the same type. 
This explicit enumeration quickly leads to path explosion in any non-trivial program, and existing approaches to addressing this problem can be roughly grouped into two categories:
\begin{itemize}
\item \emph{State merging} approaches group together the choices of {\lazy}. For instance, the work in \cite{Deng:ASE:2006,Deng:2007:TCS:1306879.1307404} represented the choices (ii) and (iii) with a variable, while the work in~\cite{Hillery:2016:EHS:2963187.2963200} captured all choices (i), (ii) and (iii) in a symbolic heap using guarded value set. Those work cannot avoid path explosion, but delay it to later stage~\cite{Deng:ASE:2006,Deng:2007:TCS:1306879.1307404}, or delegate the burden to an SMT solver~\cite{Hillery:2016:EHS:2963187.2963200}.
\item \emph{State prunning} approaches~\cite{Visser:ISSTA:2004,7004061,Braione:2015:SEP} truly mitigate the path explosion problem by using a precondition to describe \emph{some properties} of the input. After explicitly enumerating all possible choices, i.e. both valid and invalid paths, these approaches will prune the invalid paths that violate the precondition. 
\end{itemize}
A principle difference between our work and the aforementioned approaches is that we use separation logic, which is expressive enough to \emph{define} abitrary unbounded data structures. Consequently, we are able to construct valid choices from the definition, without explicit enumeration of invalid paths, and without false positives. We discuss some notable state prunning approaches in the following.
\subsubsection*{repOK} As GSE with {\lazy} results in partially initialized structures containing both concrete and symbolic values, the work in~\cite{Visser:ISSTA:2004} propose to use a particular kind of {\ok}, called \emph{conservative} or \emph{hybrid} {\ok}, that returns \form{\code{true}} when running into parts of the structure that are still symbolic. This, of course, leads to false positives.
\subsubsection*{JML} The BLISS approach in~\cite{7004061} used both hybrid {\ok} and JML~\cite{leavens1999jml} together as preconditions. The JML precondition is used to precompute relational bounds on the interpretation of class fields. It is translated into a SAT problem by the TACO tool~\cite{Galeotti:2010:AIE:1831708.1831712}. As pointed out in~\cite{Geldenhuys2013}, this translation introduces duplication, which undermines the benefit of eliminating invalid structures when the size is big. BLISS uses symmetry breaking and refine bounds to mitigate this problem.
\subsubsection*{HEX} Braione {\em et al.}~\cite{Braione:2015:SEP} introduced Heap EXploration Logic (HEX) as a specification language to constrain heap inputs. However, the language is not expressive enough to describe many common data structures, and users have to provide additional methods, called \emph{triggers}~\cite{jbse}, to check the properties that cannot be written in HEX. Moreover, HEX does not support numerical constraints, and it represents unbounded data structures using regular operators
(using \form{(\pure)^+} operator). Therefore, it is unable to capture the non-regular data structures, e.g.,
singly-linked lists which have $2^n $ nodes ($n{\geq}0$).
 Finally, it is unclear how the {HEX} solver discharges an unbounded heap formula with regular operators.
\subsubsection*{Separation logic}
Our work is also related to research on Smallfoot symbolic execution~\cite{Berdine:2005:SES:2099708.2099715} and its following work, e.g.~\cite{Chin:SCP:2012,MuellerSchwerhoffSummers16b}. Those work are not based on {\lazy}, and it is not clear how they can be used for test input generation. 
As far as we know, our work is the first to explore the use of separation logic for testing and there is only one more testing tool based on separation logic, which is a concolic execution engine named CSF~\cite{Pham:FM19}.

\section{Conclusion and Future Work} \label{sec.concl}
We present a symbolic execution framework for heap-based programs using separation logic. Our novelty is 
the proposed context-sensitive lazy initialization for test input generation.
The experimental results show that our approach significantly reduces the number of invalid test inputs and improves the test coverage.
%
For future work, we might combine JSF with bi-abduction and frame inference tools
 (i.e., Infer \cite{infer,Calcagno:JACM:2011}, S2 \cite{Le:CAV:2014,Le:TACAS:2018}) to both verify safety and
 generate test inputs
to locate/confirm real bugs in heap-based programs.
Finally, we are actively investigating the use of JSF tool for automatic program repair, a preliminary results were reported in~\cite{Zheng:2018:ADP}.\\

\noindent \textbf{Acknowledgments.}
This research is supported by the Singapore Ministry of Education
(MOE) Academic Research Fund (AcRF) Tier 1 grant.
The first author is also partially supported by the Google Summer of Code 2017 program.

\bibliographystyle{splncs04}
\bibliography{papers}

\begin{thebibliography}{10}
\providecommand{\url}[1]{\texttt{#1}}
\providecommand{\urlprefix}{URL }
\providecommand{\doi}[1]{https://doi.org/#1}

\bibitem{infer}
{Facebook Infer}. \url{https://fbinfer.com/}

\bibitem{jacoco}
{JaCoCo}. \url{https://www.eclemma.org/jacoco/}

\bibitem{jbse}
{JBSE}. \url{https://github.com/pietrobraione/jbse}

\bibitem{sir}
{SIR}. \url{http://sir.unl.edu/portal/index.php}

\bibitem{sireum}
{Sireum}. \url{https://code.google.com/archive/p/sireum/downloads}

\bibitem{sushiexperiment}
{SUSHI Experiments}. \url{https://github.com/pietrobraione/sushi-experiments}

\bibitem{Berdine:2005:SES:2099708.2099715}
Berdine, J., Calcagno, C., O'Hearn, P.W.: {Symbolic Execution with Separation
  Logic}. In: Yi, K. (ed.) APLAS 2005, pp. 52--68. Springer (2005).
  \doi{10.1007/11575467\_5}

\bibitem{Braione:ISSTA:2017}
Braione, P., Denaro, G., Mattavelli, A., Pezz\`{e}, M.: {Combining Symbolic
  Execution and Search-based Testing for Programs with Complex Heap Inputs}.
  In: Bultan, T., Sen, K. (eds.) ISSTA 2017, pp. 90--101. ACM (2017).
  \doi{10.1145/3092703.3092715}

\bibitem{Braione:2015:SEP}
Braione, P., Denaro, G., Pezz\`{e}, M.: {Symbolic Execution of Programs with
  Heap Inputs}. In: Nitto, E.D., Harman, M., Heymans, P. (eds.) FSE 2015, pp.
  602--613. ACM (2015). \doi{10.1145/2786805.2786842}

\bibitem{Braione:2016:JSE:2950290.2983940}
Braione, P., Denaro, G., Pezz\`{e}, M.: {JBSE: A Symbolic Executor for Java
  Programs with Complex Heap Inputs}. In: Zimmermann, T., Cleland{-}Huang, J.,
  Su, Z. (eds.) FSE 2016, pp. 1018--1022. ACM (2016).
  \doi{10.1145/2950290.2983940}

\bibitem{Cadar:2011:SES:1985793.1985995}
Cadar, C., Godefroid, P., Khurshid, S., P\u{a}s\u{a}reanu, C.S., Sen, K.,
  Tillmann, N., Visser, W.: {Symbolic Execution for Software Testing in
  Practice: Preliminary Assessment}. In: Taylor, R.N., Gall, H.C., Medvidovic,
  N. (eds.) ICSE 2011, pp. 1066--1071. ACM (2011).
  \doi{10.1145/1985793.1985995}

\bibitem{Calcagno:JACM:2011}
Calcagno, C., Distefano, D., O'Hearn, P.W., Yang, H.: {Compositional Shape
  Analysis by Means of Bi-Abduction}. JACM  \textbf{58}(6),  26:1--26:66
  (2011). \doi{10.1145/2049697.2049700}

\bibitem{Chin:SCP:2012}
Chin, W.N., David, C., Nguyen, H.H., Qin, S.: {Automated Verification of Shape,
  Size and Bag Properties via User-defined Predicates in Separation Logic}.
  Sci. Comput. Program.  \textbf{77}(9),  1006--1036 (2012).
  \doi{10.1016/j.scico.2010.07.004}

\bibitem{Deng:ASE:2006}
Deng, X., Lee, J., Robby: {Bogor/Kiasan: A K-bounded Symbolic Execution for
  Checking Strong Heap Properties of Open Systems}. In: ASE 2006, pp. 157--166.
  IEEE Computer Society (2006). \doi{10.1109/ASE.2006.26}

\bibitem{Deng:2007:TCS:1306879.1307404}
Deng, X., Robby, Hatcliff, J.: {Towards A Case-Optimal Symbolic Execution
  Algorithm for Analyzing Strong Properties of Object-Oriented Programs}. In:
  SEFM 2007. IEEE Computer Society (2007). \doi{10.1109/SEFM.2007.43}

\bibitem{Enderton200167}
Enderton, H.B.: {A Mathematical Introduction to Logic (Second Edition)}, pp.
  67--181. Academic Press (2001). \doi{10.1016/B978-0-08-049646-7.50008-4}

\bibitem{Galeotti:2010:AIE:1831708.1831712}
Galeotti, J.P., Rosner, N., L\'{o}pez~Pombo, C.G., Frias, M.F.: {Analysis of
  Invariants for Efficient Bounded Verification}. In: Tonella, P., Orso, A.
  (eds.) ISSTA 2010, pp. 25--36. ACM (2010). \doi{10.1145/1831708.1831712}

\bibitem{Geldenhuys2013}
Geldenhuys, J., Aguirre, N., Frias, M.F., Visser, W.: {Bounded Lazy
  Initialization}. In: Brat, G., Rungta, N., Venet, A. (eds.) NFM 2013, pp.
  229--243. Springer (2013). \doi{10.1007/978-3-642-38088-4\_16}

\bibitem{Hillery:2016:EHS:2963187.2963200}
Hillery, B., Mercer, E., Rungta, N., Person, S.: {Exact Heap Summaries for
  Symbolic Execution}. In: Jobstmann, B., Leino, K.R.M. (eds.) VMCAI 2016, pp.
  206--225. Springer (2016). \doi{10.1007/978-3-662-49122-5\_10}

\bibitem{Ishtiaq:2001:BAL:360204.375719}
Ishtiaq, S.S., O'Hearn, P.W.: {BI as an Assertion Language for Mutable Data
  Structures}. In: Hankin, C., Schmidt, D. (eds.) POPL 2001, pp. 14--26. ACM
  (2001). \doi{10.1145/360204.375719}

\bibitem{Khurshid2003}
Khurshid, S., P\u{a}s\u{a}reanu, C.S., Visser, W.: {Generalized Symbolic
  Execution for Model Checking and Testing}. In: Garavel, H., Hatcliff, J.
  (eds.) TACAS 2003, pp. 553--568. Springer (2003).
  \doi{10.1007/3-540-36577-X\_40}

\bibitem{King:1976:SEP:360248.360252}
King, J.C.: {Symbolic Execution and Program Testing}. Commun. ACM
  \textbf{19}(7),  385--394 (1976). \doi{10.1145/360248.360252}

\bibitem{Le:CAV:2014}
Le, Q.L., Gherghina, C., Qin, S., Chin, W.N.: {Shape Analysis via Second-Order
  Bi-Abduction}. In: Biere, A., Bloem, R. (eds.) CAV 2014, pp. 52--68. Springer
  (2014). \doi{10.1007/978-3-319-08867-9\_4}

\bibitem{Le:CAV:2016}
Le, Q.L., Sun, J., Chin, W.N.: {Satisfiability Modulo Heap-Based Programs}. In:
  Chaudhuri, S., Farzan, A. (eds.) CAV 2016, pp. 382--404. Springer (2016).
  \doi{10.1007/978-3-319-41528-4\_21}

\bibitem{Le:TACAS:2018}
Le, Q.L., Sun, J., Qin, S.: {Frame Inference for Inductive Entailment Proofs in
  Separation Logic}. In: Beyer, D., Huisman, M. (eds.) {TACAS 2018}, pp.
  41--60. Springer (2018). \doi{10.1007/978-3-319-89960-2\_3}

\bibitem{DBLP:conf/cav/LeT0C17}
Le, Q.L., Tatsuta, M., Sun, J., Chin, W.: {A Decidable Fragment in Separation
  Logic with Inductive Predicates and Arithmetic}. In: Majumdar, R., Kuncak, V.
  (eds.) CAV 2017, pp. 495--517. Springer (2017).
  \doi{10.1007/978-3-319-63390-9\_26}

\bibitem{leavens1999jml}
Leavens, G.T., Baker, A.L., Ruby, C.: {JML: A Notation for Detailed Design}.
  In: Kilov, H., Rumpe, B., Simmonds, I. (eds.) Behavioral Specifications of
  Businesses and Systems, pp. 175--188. Springer (1999).
  \doi{10.1007/978-1-4615-5229-1\_12}

\bibitem{MuellerSchwerhoffSummers16b}
M{\"{u}}ller, P., Schwerhoff, M., Summers, A.J.: {Automatic Verification of
  Iterated Separating Conjunctions using Symbolic Execution}. In: Chaudhuri,
  S., Farzan, A. (eds.) CAV 2016, pp. 405--425. Springer (2016).
  \doi{10.1007/978-3-319-41528-4\_22}

\bibitem{Pham:FM19}
Pham, L.H., Le, Q.L., Phan, Q.S., Sun, J.: {Concolic Testing Heap-Manipulating
  Programs}. In: FM 2019. To appear

\bibitem{Pham:2018:THP:3183440.3194964}
Pham, L.H., Le, Q.L., Phan, Q.S., Sun, J., Qin, S.: {Testing Heap-based
  Programs with Java StarFinder}. In: Chaudron, M., Crnkovic, I., Chechik, M.,
  Harman, M. (eds.) ICSE 2018, pp. 268--269. ACM (2018).
  \doi{10.1145/3183440.3194964}

\bibitem{Pasareanu:2013:ASE}
P\u{a}s\u{a}reanu, C.S., Visser, W., Bushnell, D., Geldenhuys, J., Mehlitz, P.,
  Rungta, N.: {Symbolic PathFinder: Integrating Symbolic Execution with Model
  Checking for Java Bytecode Analysis}. Autom. Softw. Eng.  \textbf{20}(3),
  391--425 (2013). \doi{10.1007/s10515-013-0122-2}

\bibitem{Reynolds:LICS02}
Reynolds, J.: {Separation Logic: A Logic for Shared Mutable Data Structures}.
  In: LICS 2002, pp. 55--74. {IEEE} Computer Society (2002).
  \doi{10.1109/LICS.2002.1029817}

\bibitem{7004061}
Rosner, N., Geldenhuys, J., Aguirre, N., Visser, W., Frias, M.F.: {BLISS:
  Improved Symbolic Execution by Bounded Lazy Initialization with SAT Support}.
  {IEEE} Trans. Software Eng.  \textbf{41}(7),  639--660 (2015).
  \doi{10.1109/TSE.2015.2389225}

\bibitem{Schwartz:2010:YEW:1849417.1849981}
Schwartz, E.J., Avgerinos, T., Brumley, D.: {All You Ever Wanted to Know about
  Dynamic Taint Analysis and Forward Symbolic Execution (but Might Have Been
  Afraid to Ask)}. In: S{\&}P 2010, pp. 317--331. IEEE Computer Society (2010).
  \doi{10.1109/SP.2010.26}

\bibitem{Tatsuta:APLAS:2016}
Tatsuta, M., Le, Q.L., Chin, W.N.: {Decision Procedure for Separation Logic
  with Inductive Definitions and Presburger Arithmetic}. In: Igarashi, A. (ed.)
  APLAS 2016, pp. 423--443. Springer (2016).
  \doi{10.1007/978-3-319-47958-3\_22}

\bibitem{Visser:ISSTA:2004}
Visser, W., P\v{a}s\v{a}reanu, C.S., Khurshid, S.: {Test Input Generation with
  Java PathFinder}. In: Avrunin, G.S., Rothermel, G. (eds.) ISSTA 2004, pp.
  97--107. ACM (2004). \doi{10.1145/1007512.1007526}

\bibitem{Zheng:2018:ADP}
Zheng, G., Le, Q.L., Nguyen, T., Phan, Q.S.: {Automatic Data Structure Repair
  using Separation Logic}. In: {JPF 2018}

\end{thebibliography}
\end{document}